\documentclass{article}
\usepackage{biblatex}
\usepackage[margin=1in]{geometry}
\usepackage{amsmath}
\usepackage{amssymb}
\usepackage[hidelinks]{hyperref}
\usepackage{graphicx}
\usepackage{longtable}
\usepackage[ruled,vlined]{algorithm2e}
\usepackage{booktabs}

\hypersetup{
  pdftitle={New lower bounds for constant-weight codes via seeded bit-swap tabu search},
  pdfauthor={William Echols}
}

\title{New lower bounds for constant-weight codes via seeded bit-swap tabu search}
\author{William Echols}
\date{}

\begin{document}

\maketitle

\begin{abstract}
A binary constant-weight code is a set of binary words of length $n$ such that each word has exactly weight $w$ and is at least Hamming distance $d$ from every other word in the set. $A(n,d,w)$ denotes the maximum size of a binary constant-weight code with parameters $(n,d,w)$. Using seeded initialization with bit-swap tabu search, we found 124 new constructions that improve existing lower bounds for $A(n,d,w)$. As a corollary of stronger bounds on $A(n,8,8)$ for $n \in \{ 32,33,34,37 \}$, we also improve lower bounds on kissing numbers $\tau_{32}$, $\tau_{33}$, $\tau_{34}$, and $\tau_{37}$.
\end{abstract}

\section{Introduction}

A \textit{binary constant-weight code} is a set of binary words, where each word has length $n$, Hamming weight $w$, and is at least Hamming distance $d$ from every other word in the set. The Hamming distance of two binary words of length $n$, denoted $x$ and $y$, is defined to be $d_H(x,y) = \sum_{i=1}^n(x_i \oplus y_i)$, and the Hamming weight of a given binary word $x$ is defined as $w_H(x) = \sum_{i=1}^n x_i$. The maximum number of binary codewords of length $n$ and Hamming weight $w$ such that any two codewords are at least Hamming distance $d$ apart is denoted as $A(n,d,w)$.

Significant computational efforts have been used to construct codes for various choices of $(n,d,w)$, demonstrating lower bounds on $A(n,d,w)$ \cite{brouwer1990, montemanni2009, smith2006, smith2012}. In addition, existing codes for parameter set $(n,d,w)$ can be lengthened to find codes for $(n+1,d,w)$ or shortened to find codes for $(n-1,d,w)$ \cite{brouwer1990, chee2010}. We focus on initialization via seeding, from both equivalent and neighboring parameters, to find stronger lower bounds for $A(n,d,w)$.

Previous implementations of tabu search for the construction of constant-weight codes resulted in weaker bounds than known results and alternative metaheuristic algorithms \cite{bland1997, montemanni2009}. However, the formulation of a bit-swap tabu search by Rosin found new lower bounds for six constant-weight codes \cite{rosin2026}. 

\section{Methods}

We modify Rosin's approach \cite{rosin2026} by warm-starting bit-swap tabu search; rather than initializing the search with $s$ randomly generated $n$-bit words of weight $w$, we instead seed the search using codewords from an existing code for the same or neighboring parameter set. 

In the direct case, where the target has parameters $(n,d,w)$ and seeding uses an existing code $S$ of the same parameters and word count $m$, we first construct a library composed of the existing codewords appended with an additional word $v$ chosen to minimize the aggregate distance-deficit penalty; we use a library rather than a single candidate code as there are multiple words with equivalent minimal-deficit penalties. Then, Rosin's refined bit-swap tabu search is executed up to $4|E|$ times with random samples from the library, where $E$ is the set of minimal-deficit words. We use Rosin's tabu parameters unchanged \cite{rosin2026} but omit the restart mechanism which would otherwise override the provided seed. Let $W(n,w) = \{ x\in \{0,1\}^n : \sum_{i=1}^n x_i = w \}$ be the set of binary words of length $n$ and Hamming weight $w$. Direct seeded initialization is described in Algorithm \ref{alg:direct_seed_initialization}.

We broaden the seeding concept to allow the strengthening of certificates with adjacent values of $n$, as described in Algorithm \ref{alg:neighbor_seeded_initialization}. Neighbor-seeded initialization includes code lengthening, which uses an existing certificate for $A(n-1,d,w)$ to form words of length $n$; these words then constitute the initial code for the bit-swap tabu search. Symmetrically, a certificate for $A(n+1,d,w)$ can be shortened on a zero-slice to seed a search for parameter set $(n,d,w)$. The intermediate code obtained by lengthening or shortening is usually smaller than the target size $s$; we pad it to size $s$ by appending randomly chosen words of minimum penalty, at the cost of violating the minimum-distance property. Recomputing the minimal-deficit set $E$ after each insertion is computationally intensive; as a result, our implementation computes $E$ initially and samples from it uniformly with replacement. The bit-swap tabu search then repairs the padded set into a valid code. Both the lengthening and shortening techniques are well-known in the literature \cite{brouwer1990, chee2010} and have been used to construct lower bounds \cite{brouwer}.

Since each successful run produces a new valid code, the algorithm can be applied successively to its own output. Varying the seeding method between iterations can additionally allow the search to escape local obstacles through the use of words from neighboring parameter sets.

\section{Results}

Table~\ref{tab:new_lower_bounds} lists 124 new constructions which were discovered with these methods. Each code improves a lower bound in Brouwer's online table \cite{brouwer} and was verified with a separate program in the repository.

\subsection{Constant-weight codes}

\begingroup\setlength{\tabcolsep}{5pt}
\begin{longtable}{@{}r r r r r r @{\hspace{6em}} r r r r r r@{}}
\caption{New lower bounds for $A(n,d,w)$. Prior lower bounds are sourced from Brouwer's online table \cite{brouwer}.}
\label{tab:new_lower_bounds}
\\
$n$ & $d$ & $w$ & Prior & \textbf{New} & Gain
&
$n$ & $d$ & $w$ & Prior & \textbf{New} & Gain
\\
\cmidrule(r{6em}){1-6}\cmidrule{7-12}
\endfirsthead
\multicolumn{12}{c}{\tablename\ \thetable{} -- continued}
\\
\cmidrule(r{6em}){1-6}\cmidrule{7-12}
$n$ & $d$ & $w$ & Prior & \textbf{New} & Gain
&
$n$ & $d$ & $w$ & Prior & \textbf{New} & Gain
\\
\cmidrule(r{6em}){1-6}\cmidrule{7-12}
\endhead
\cmidrule(r{6em}){1-6}\cmidrule{7-12}
\multicolumn{12}{r}{\textit{Continued on the next page}}
\\
\endfoot
\cmidrule(r{6em}){1-6}\cmidrule{7-12}
\endlastfoot
21 & 6 & 5 & 108 & \textbf{109} & +1 & 34 & 10 & 7 & 75 & \textbf{85} & +10 \\
43 & 6 & 5 & 1077 & \textbf{1081} & +4 & 35 & 10 & 7 & 86 & \textbf{102} & +16 \\
44 & 6 & 5 & 1131 & \textbf{1132} & +1 & 36 & 10 & 7 & 103 & \textbf{116} & +13 \\
54 & 6 & 5 & 2148 & \textbf{2157} & +9 & 37 & 10 & 7 & 122 & \textbf{134} & +12 \\
55 & 6 & 5 & 2355 & \textbf{2365} & +10 & 38 & 10 & 7 & 138 & \textbf{152} & +14 \\
56 & 6 & 5 & 2414 & \textbf{2477} & +63 & 39 & 10 & 7 & 156 & \textbf{170} & +14 \\
58 & 6 & 5 & 2639 & \textbf{2677} & +38 & 40 & 10 & 7 & 178 & \textbf{189} & +11 \\
59 & 6 & 5 & 2869 & \textbf{2870} & +1 & 41 & 10 & 7 & 198 & \textbf{212} & +14 \\
 &  &  &  &  &  & 42 & 10 & 7 & 220 & \textbf{233} & +13 \\
30 & 6 & 6 & 1277 & \textbf{1316} & +39 & 43 & 10 & 7 & 240 & \textbf{250} & +10 \\
34 & 6 & 6 & 2003 & \textbf{2017} & +14 & 44 & 10 & 7 & 264 & \textbf{273} & +9 \\
35 & 6 & 6 & 2187 & \textbf{2204} & +17 & 45 & 10 & 7 & 290 & \textbf{293} & +3 \\
 &  &  &  &  &  & 46 & 10 & 7 & 311 & \textbf{314} & +3 \\
20 & 6 & 8 & 588 & \textbf{590} & +2 & 50 & 10 & 7 & 396 & \textbf{402} & +6 \\
22 & 6 & 8 & 1144 & \textbf{1145} & +1 & 51 & 10 & 7 & 406 & \textbf{411} & +5 \\
 &  &  &  &  &  & 52 & 10 & 7 & 411 & \textbf{420} & +9 \\
28 & 8 & 6 & 131 & \textbf{132} & +1 & 53 & 10 & 7 & 420 & \textbf{428} & +8 \\
38 & 8 & 6 & 302 & \textbf{303} & +1 & 55 & 10 & 7 & 514 & \textbf{519} & +5 \\
42 & 8 & 6 & 441 & \textbf{448} & +7 & 56 & 10 & 7 & 583 & \textbf{586} & +3 \\
43 & 8 & 6 & 454 & \textbf{457} & +3 & 57 & 10 & 7 & 589 & \textbf{600} & +11 \\
44 & 8 & 6 & 468 & \textbf{470} & +2 & 58 & 10 & 7 & 602 & \textbf{612} & +10 \\
45 & 8 & 6 & 489 & \textbf{491} & +2 & 59 & 10 & 7 & 611 & \textbf{622} & +11 \\
48 & 8 & 6 & 663 & \textbf{670} & +7 & 60 & 10 & 7 & 621 & \textbf{638} & +17 \\
49 & 8 & 6 & 683 & \textbf{685} & +2 & 61 & 10 & 7 & 668 & \textbf{675} & +7 \\
56 & 8 & 6 & 1162 & \textbf{1163} & +1 & 62 & 10 & 7 & 746 & \textbf{749} & +3 \\
 &  &  &  &  &  & 63 & 10 & 7 & 831 & \textbf{835} & +4 \\
26 & 8 & 7 & 259 & \textbf{264} & +5 &  &  &  &  &  &  \\
28 & 8 & 7 & 318 & \textbf{321} & +3 & 24 & 10 & 8 & 38 & \textbf{39} & +1 \\
29 & 8 & 7 & 344 & \textbf{351} & +7 & 26 & 10 & 8 & 55 & \textbf{68} & +13 \\
30 & 8 & 7 & 389 & \textbf{404} & +15 & 27 & 10 & 8 & 66 & \textbf{76} & +10 \\
31 & 8 & 7 & 463 & \textbf{475} & +12 & 28 & 10 & 8 & 81 & \textbf{85} & +4 \\
32 & 8 & 7 & 500 & \textbf{514} & +14 & 29 & 10 & 8 & 91 & \textbf{96} & +5 \\
33 & 8 & 7 & 539 & \textbf{557} & +18 & 36 & 10 & 8 & 216 & \textbf{245} & +29 \\
34 & 8 & 7 & 594 & \textbf{600} & +6 & 43 & 10 & 8 & 456 & \textbf{470} & +14 \\
36 & 8 & 7 & 730 & \textbf{754} & +24 & 47 & 10 & 8 & 793 & \textbf{795} & +2 \\
37 & 8 & 7 & 851 & \textbf{873} & +22 & 48 & 10 & 8 & 896 & \textbf{904} & +8 \\
38 & 8 & 7 & 932 & \textbf{974} & +42 & 49 & 10 & 8 & 1029 & \textbf{1035} & +6 \\
39 & 8 & 7 & 1014 & \textbf{1118} & +104 & 50 & 10 & 8 & 1172 & \textbf{1191} & +19 \\
40 & 8 & 7 & 1170 & \textbf{1230} & +60 & 51 & 10 & 8 & 1358 & \textbf{1370} & +12 \\
41 & 8 & 7 & 1287 & \textbf{1293} & +6 & 52 & 10 & 8 & 1565 & \textbf{1572} & +7 \\
45 & 8 & 7 & 1867 & \textbf{1885} & +18 &  &  &  &  &  &  \\
52 & 8 & 7 & 3795 & \textbf{3796} & +1 & 32 & 10 & 9 & 304 & \textbf{305} & +1 \\
53 & 8 & 7 & 3978 & \textbf{3979} & +1 &  &  &  &  &  &  \\
54 & 8 & 7 & 4168 & \textbf{4172} & +4 & 30 & 10 & 11 & 504 & \textbf{512} & +8 \\
55 & 8 & 7 & 4337 & \textbf{4386} & +49 &  &  &  &  &  &  \\
 &  &  &  &  &  & 25 & 10 & 12 & 137 & \textbf{141} & +4 \\
27 & 8 & 8 & 769 & \textbf{775} & +6 &  &  &  &  &  &  \\
30 & 8 & 8 & 1145 & \textbf{1162} & +17 & 26 & 10 & 13 & 225 & \textbf{234} & +9 \\
31 & 8 & 8 & 1330 & \textbf{1352} & +22 & 27 & 10 & 13 & 420 & \textbf{425} & +5 \\
32 & 8 & 8 & 1659 & \textbf{1667} & +8 &  &  &  &  &  &  \\
33 & 8 & 8 & 1777 & \textbf{1788} & +11 & 28 & 10 & 14 & 812 & \textbf{814} & +2 \\
34 & 8 & 8 & 1934 & \textbf{1936} & +2 & 31 & 10 & 14 & 1538 & \textbf{1539} & +1 \\
37 & 8 & 8 & 2817 & \textbf{2832} & +15 &  &  &  &  &  &  \\
38 & 8 & 8 & 2997 & \textbf{3025} & +28 & 34 & 12 & 8 & 38 & \textbf{39} & +1 \\
39 & 8 & 8 & 3323 & \textbf{3324} & +1 & 35 & 12 & 8 & 40 & \textbf{44} & +4 \\
 &  &  &  &  &  & 36 & 12 & 8 & 48 & \textbf{50} & +2 \\
26 & 8 & 9 & 887 & \textbf{890} & +3 & 37 & 12 & 8 & 49 & \textbf{53} & +4 \\
27 & 8 & 9 & 1025 & \textbf{1029} & +4 & 38 & 12 & 8 & 57 & \textbf{60} & +3 \\
28 & 8 & 9 & 1333 & \textbf{1375} & +42 & 56 & 12 & 8 & 362 & \textbf{371} & +9 \\
 &  &  &  &  &  & 57 & 12 & 8 & 367 & \textbf{374} & +7 \\
27 & 8 & 10 & 1721 & \textbf{1727} & +6 & 58 & 12 & 8 & 384 & \textbf{401} & +17 \\
28 & 8 & 10 & 2067 & \textbf{2103} & +36 & 59 & 12 & 8 & 419 & \textbf{425} & +6 \\
 &  &  &  &  &  & 62 & 12 & 8 & 495 & \textbf{497} & +2 \\
25 & 8 & 11 & 1702 & \textbf{1722} & +20 &  &  &  &  &  &  \\
26 & 8 & 11 & 2037 & \textbf{2042} & +5 & 29 & 12 & 10 & 66 & \textbf{70} & +4 \\
 &  &  &  &  &  & 30 & 12 & 10 & 98 & \textbf{101} & +3 \\
27 & 8 & 12 & 3335 & \textbf{3337} & +2 & 31 & 12 & 10 & 103 & \textbf{108} & +5 \\
\\
 &  &  &  &  &  & 29 & 12 & 13 & 146 & \textbf{153} & +7 \\
 &  &  &  &  &  & 30 & 12 & 13 & 236 & \textbf{239} & +3 \\
\\
 &  &  &  &  &  & 29 & 12 & 14 & 173 & \textbf{174} & +1 \\
\\
 &  &  &  &  &  & 29 & 14 & 13 & 35 & \textbf{38} & +3 \\
 &  &  &  &  &  & 30 & 14 & 13 & 45 & \textbf{46} & +1 \\
\\
 &  &  &  &  &  & 34 & 14 & 14 & 131 & \textbf{132} & +1 \\
 &  &  &  &  &  & 38 & 14 & 14 & 342 & \textbf{343} & +1 \\
\end{longtable}
\endgroup

As an immediate corollary of the codes in Table \ref{tab:new_lower_bounds}, appending a zero coordinate to our $(28,8,6)$ code gives $A(29,8,6) \ge 132$, and shortening our $(48,8,6)$ code on a zero-slice gives $A(47,8,6) \ge 588$. Each of these improves the previously listed bounds of 130 and 583, respectively \cite{brouwer}.

\subsection{Kissing numbers}

$A(n, d)$ denotes the maximum size of a binary code such that all words have length $n$ and any two words are at least Hamming distance $d$ apart. In particular, the $[32,17,8]$ binary linear code of Cheng and Sloane gives $A(32,8) \ge 2^{17}$ \cite{cheng1989}.

From Edel et al. \cite{edel1998}, we have the following lower bound on the kissing number $\tau_n$, where the chosen values of $n_0,n_1,\ldots,n_\mu \in \mathbb{N}$ satisfy $n\ge n_0$ and $\forall i \in \{ 0,1,\ldots,\mu-1 \},n_i \ge 4n_{i+1} \ge 1$.

\begin{align}
\label{kissing_number_lower_bound}
\tau_n &\ge \sum_{\nu = 0}^\mu A(n,n_\nu,n_\nu)A\left(n_\nu,\left\lceil \frac{n_\nu}{4} \right\rceil\right)
\end{align}

With choices $(n,n_0,n_1,n_2) = (32, 32, 8, 2)$ and the previously-known lower bound $A(32,8,8) \ge 1659$, we find

\begin{align*}
\tau_{32} &\ge A(32,32,32)A(32,8) + A(32,8,8)A(8,2) + A(32,2,2)A(2,1) \\
&\ge 1 \cdot 2^{17} + 1659 \cdot 2^7 + 496 \cdot 2^2 \\
&\ge 345408
\end{align*}

With the improvement of $A(32,8,8) \ge 1667$, we can immediately strengthen the bound to $\tau_{32} \ge 346432$. For $n \ge 32$, we can modify Equation \eqref{kissing_number_lower_bound} with choices $(n_0, n_1, n_2) = (32, 8, 2)$ to show

\begin{align}
\tau_{n} \ge 2^{17} + A(n,8,8) \cdot 2^7 + 2n(n-1)
\end{align}

The new lower bounds along with their associated choices for $n_0,n_1\ldots,n_\mu$ are presented in Table \ref{tab:kissing_lower_bounds}.

\begin{center}
\small
\begin{longtable}{ c c c c c }
\caption{New lower bounds for $\tau_n$. Derived from improved lower bounds of $A(n,8,8)$. Prior lower bounds sourced from Brouwer \cite{brouwer}.}
\label{tab:kissing_lower_bounds} \\

\hline
$\tau_n$ & $n_0,n_1,\ldots,n_\mu$ & Prior & \textbf{New} & Gain \\
\hline
\endfirsthead

\multicolumn{5}{c}{\tablename\ \thetable{} -- continued} \\
\hline
$\tau_n$ & $n_0,n_1,\ldots,n_\mu$ & Prior & \textbf{New} & Gain \\
\hline
\endhead

\hline
\multicolumn{5}{r}{\textit{Continued on next page}} \\
\endfoot

\hline
\endlastfoot
$\tau_{32}$ & 32, 8, 2 & 345408 & \textbf{346432} & +1024 \\
$\tau_{33}$ & 32, 8, 2 & 360640 & \textbf{362048} & +1408 \\
$\tau_{34}$ & 32, 8, 2 & 380868 & \textbf{381124} & +256 \\
$\tau_{37}$ & 32, 8, 2 & 494312 & \textbf{496232} & +1920
\end{longtable}
\end{center}

\section*{Acknowledgments}

The author thanks Christopher D. Rosin for his verification of an initial result and helpful correspondence. The author also thanks Andries Brouwer for the maintenance of the online table \cite{brouwer} and verification of codes.

\section*{Data Availability}

The binary constant-weight codes referenced in Table \ref{tab:new_lower_bounds} are publicly available at \url{https://github.com/williamechols/cwc-seeded-tabu-search}.

\clearpage
\appendix

\section{Algorithms}

\begin{algorithm}[H]
\DontPrintSemicolon
\caption{Direct initialization}
\label{alg:direct_seed_initialization}
\KwIn{$(n,d,w)$ code $S$ of size $m$}
\KwOut{Library $\mathcal{L}$ of initialization sets with $m+1$ words of length $n$ and weight $w$}
$E \gets \arg\min_{x\in W(n,w)\setminus S} \sum_{s\in S}\max(0,d-d_H(x,s))$\;
$\mathcal{L} \gets \{ S \cup \{ e \} : e \in E \}$\;
\Return{$\mathcal{L}$}
\end{algorithm}

\begin{algorithm}[H]
\DontPrintSemicolon
\caption{Neighbor initialization}
\label{alg:neighbor_seeded_initialization}
\KwIn{Parameters $(n,d,w)$, target size $s$, and library $\mathcal{L}$ of existing constant-weight codes}
\KwOut{Library $\mathcal{P}$ of initialization sets with $s$ words of length $n$ and weight $w$}
\SetKwProg{Fn}{Function}{:}{}
\SetKwFunction{Pad}{PadToSize}
\BlankLine
\Fn{\Pad{$C,s$}}{
    $X \gets \text{ordered list of }C$\;
    \If{$|X| \ge s$}{
        \KwRet{the first $s$ elements of $X$}\;
    }
    $E \gets \arg\min\limits_{x\in W(n,w)\setminus X}\sum_{z\in X}\max(0,d-d_H(x,z))$\;
    \While{$|X|<s$}{
        $e \gets \text{random selection from }E$\;
        $\text{Append }e\text{ to }X$\;
    }
    \KwRet{$X$}\;
}
\BlankLine
$\mathcal{S}\gets\varnothing$\;
$\mathcal{P}\gets\varnothing$\;
\BlankLine
\tcp{\textbf{Direct shortening seed}}
$B \gets \text{the largest }(n+1,d,w)\text{ code in }\mathcal{L}$\;
\If{$B$ exists}{
    \For{$j \gets 1$ \KwTo $n+1$}{
        $C_j \gets \{x\text{ without coordinate }j : x\in B,\ x_j=0\}$\;
    }
    $C_{\mathrm{short}} \gets \arg\max\limits_{1\le j\le n+1}|C_j|$\;
    $\mathcal{S} \gets \mathcal{S} \cup \{C_{\mathrm{short}}\}$\;
}
\BlankLine
\tcp{\textbf{Zero-slice lengthening seed}}
$A \gets \text{the largest }(n-1,d,w)\text{ code in }\mathcal{L}$\;
\If{$A$ exists}{
    $C_0 \gets \{0x : x \in A\}$\;
    $F \gets \{ y \in \{0,1\}^{n-1} : w_H(y)=w-1
    \text{ and } \forall x\in A,\ d_H(x,y)\ge d-1 \}$\;
    $Q \gets \varnothing$\;
    \While{$|C_0|+|Q|<s$}{
        $F' \gets \{u\in F : \forall v\in Q,\ d_H(u,v)\ge d\}$\;
        \If{$F'=\varnothing$}{
            \textbf{break}\;
        }
        $u \gets \text{random selection from }F'$\;
        $Q \gets Q\cup\{u\}$\;
        $F \gets F\setminus\{u\}$\;
    }
    $C_{\mathrm{len}} \gets C_0 \cup \{1y : y\in Q\}$\;
    $\mathcal{S} \gets \mathcal{S} \cup \{C_{\mathrm{len}}\}$\;
}
\BlankLine
\ForEach{$C\in\mathcal{S}$}{
    $\mathcal{P}\gets\mathcal{P}\cup\{\Pad(C,s)\}$\;
}
\KwRet{$\mathcal{P}$}
\end{algorithm}

\printbibliography

\end{document}